\documentclass[twocolumn,showpacs,preprintnumbers,amsmath,amssymb,floatfix]{revtex4}
\usepackage{graphicx}% Include figure files
\usepackage{dcolumn}% Align table columns on decimal point
\usepackage{bm}% bold math
\usepackage{color}
\begin{document}

\title{
Magnon Supersolid and Anomalous Hysteresis in Spin Dimers on a Triangular Lattice
}

\author{Daisuke Yamamoto$^{1}$}
\author{Ippei Danshita$^{2,3}$}
\affiliation{
{$^1$Condensed Matter Theory Laboratory, RIKEN, Wako, Saitama 351-0198, Japan}
\\
{$^2$Yukawa Institute for Theoretical Physics, Kyoto University, Kyoto 606-8502, Japan}
\\
{$^3$Computational Condensed Matter Physics Laboratory, RIKEN, Wako, Saitama 351-0198, Japan}
}
\date{\today}% It is always \today, today,
             % but any date may be explicitly specified
\begin{abstract}
We study the magnetic phase diagram and hysteresis behavior of weakly coupled spin dimers on a triangular lattice using the cluster mean-field method with cluster-size scaling. We find that the magnetization curve has plateaus at 1/3 and 2/3 of the total magnetization, in which local singlet and triplet states form a superlattice pattern. Moreover, if increasing (decreasing) the magnetic field from the 1/3 (2/3) plateau, the Bose-Einstein condensation (BEC) of triplons occurs on the superlattice background, leading to the transition into {\it magnon supersolid} phase. We also find that the first-order transition between these solid states and the standard magnon BEC state exhibits an anomalous hysteresis upon cycling the magnetic field; the transition can occur only from solid to BEC, and the system cannot return to the initial solid state in the reverse process. 
\end{abstract}
\maketitle

%%%%%%%%%%%%%%%%%%%%%%%%%%%%%%%%%%%%%%%%%%%%%%%%%%%%%%%%%%%%%%%%%%%%%%%%%%%%%%%
%%%%%%%%%%%%%%%%%%%%%%%%%%%%%%%%%%%%%%%%%%%%%%%%%%%%%%%%%%%%%%%%%%%%%%%%%%%%%%%
%%                                                                           %%
%% Section I: introduction                                                   %%
%%                                                                           %%
%%%%%%%%%%%%%%%%%%%%%%%%%%%%%%%%%%%%%%%%%%%%%%%%%%%%%%%%%%%%%%%%%%%%%%%%%%%%%%%
%%%%%%%%%%%%%%%%%%%%%%%%%%%%%%%%%%%%%%%%%%%%%%%%%%%%%%%%%%%%%%%%%%%%%%%%%%%%%%%
\section{\label{1}Introduction} 
% 1st
The physics of bosonic elementary excitations in coupled spin-dimer systems has been attracting considerable attention since the observation of magnon Bose-Einstein condensation (BEC) in TlCuCl$_3$~\cite{oosawa-99,nikuni-00,giamarchi-08}. The strong intradimer antiferromagnetic interaction in spin-dimer systems~\cite{jaime-04,uchida-01,stone-08,samulon-09,conner-10} provides a quantum disordered (QD) singlet state, and the system undergoes a phase transition to a magnetically ordered state when applying a sufficiently strong magnetic field. This quantum phase transition is now well established as a BEC of spin-triplet states with $S^z=+1$, called ``triplons,'' moving in the nonmagnetic singlet background~\cite{nikuni-00,giamarchi-08}. Moreover, some materials exhibit magnetization plateaus at fractional values of the total magnetization~\cite{kageyama-99,kodama-02}, where the triplons are crystallized in a superstructure pattern and form an incompressible ``solid'' state. As a next step, it is naturally expected that one could observe a more exotic state, namely $the$ $magnon$ $supersolid$ $(SS)$~\cite{chen-10,sengupta-07,albuquerque-11,ng-06,laflorencie-07,picon-08,momoi-00,murakami-13}, in which the diagonal (solid) and off-diagonal (magnon BEC) long-range orders coexist~\cite{matsuda-70}. 

% 2nd
The possibility of SS states has been discussed also in the contexts of solid helium-4~\cite{Ekim-04,day-07,maris-12,DYkim-12} and ultracold Bose gases in optical lattices~\cite{goral-02,sengupta-05,wessel-05,danshita-09,sansone-10,pollet-10,ohgoe-11,yamamoto-12}.
In the former case, although Kim and Chan's torsional oscillator experiments observed superficial nonclassical rotational inertia in 2004~\cite{Ekim-04}, it is now interpreted as shear modulus stiffening rather than superfluidity in the latest experiments with an improved oscillator~\cite{DYkim-12}.
In the latter case, previous theoretical studies have established that sufficiently strong dipole-dipole interactions between atoms or molecules, which are long-ranged, allow for the emergence of SS phases~\cite{goral-02,sengupta-05,wessel-05,danshita-09,sansone-10,pollet-10,ohgoe-11,yamamoto-12}. 
However, the strength of dipolar interactions available with current technology is still significantly smaller than experimentally achievable temperatures, and SS has not been realized yet.
Given the presence of a rich variety of materials and the large spin-exchange interactions, possibly allowing SS states to be realized at achievable
temperatures, spin-dimer systems are advantageous over the other
candidate SS systems.

\begin{figure}[b]
\includegraphics[scale=0.375]{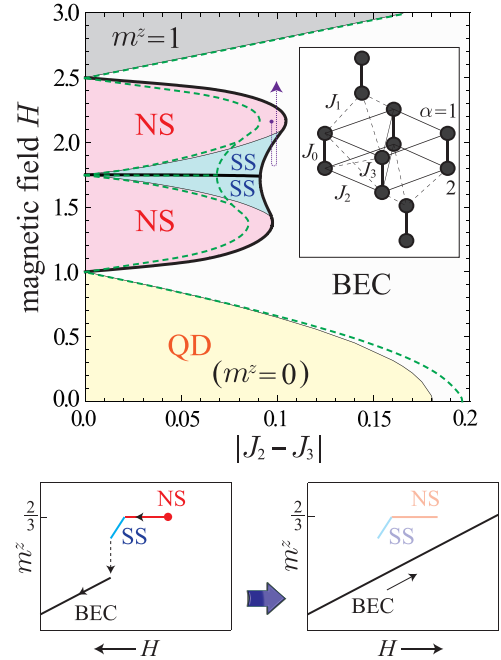}
\caption{\label{fig1} (Color online)
Ground-state phase diagram of spin-$1/2$ dimers on the triangular lattice obtained by the CMF-9 calculations. We set $J_0 = 2 (J_2 + J_3) = 1$. The dashed green lines are the phase boundaries determined by the cluster-size scaling (CMF+S). Lower panels: Schema of the anomalous hysteresis process when decreasing and then increasing the magnetic field from an initial NS state along the route marked by the purple arrow in the phase diagram. 
}
\end{figure}
%%
%

% 3rd
Nevertheless, no clear SS behavior has been observed in any spin-dimer materials so far. 
This may be because the parameter window for SS is very small.
Previous theoretical analyses with square-lattice models~\cite{chen-10,sengupta-07,albuquerque-11,ng-06,laflorencie-07,picon-08} have indeed shown that the magnon SS state can emerge only in a very narrow range of parameters (e.g., magnetic field) even when it exists.
In this paper, we study the ground-state properties and hysteresis behaviors of weakly coupled spin-dimers on a triangular lattice. Taking into consideration that in the case of lattice-boson systems the SS states on a triangular lattice are fairly robust against a domain-wall formation unlike in the square-lattice case~\cite{sengupta-05,wessel-05,ohgoe-11}, it is speculated that triangular-lattice geometry is more suitable for exploring SS states also in spin-dimer systems.
Furthermore, a novel anomalous hysteresis, in which the transition between two phases occurs only unidirectionally, has been predicted in lattice-boson systems~\cite{yamamoto-12}. Such a unidirectional behavior without forming a ``hysteresis loop'' has not been observed in experiments and has not even been predicted theoretically for magnetic hysteresis in any materials. Therefore, it is imperative to examine whether spin dimers can exhibit such anomalous hysteresis.

% 4th
We quantitatively determine the ground-state phase diagram of the triangular-lattice spin-dimer model even far away from the strongly dimerized limit as shown in Fig.~\ref{fig1}. 
The frustration coming from the competition of two interdimer interactions makes numerical analyses such as quantum Monte Carlo simulations rather difficult and time consuming for frustrated spin-dimer systems. Because of this problem, there has been no theoretical (quantitatively reliable) prediction of the complete phase diagram for frustrated spin-dimer models even in the square-lattice case as far as we know~\cite{sengupta-07,chen-10,albuquerque-11}. To overcome this difficulty, we use a large-size cluster mean-field theory with a scaling scheme~(CMF+S)~\cite{yamamoto-12-2}. As shown in Fig.~\ref{fig1}, the magnon SS state emerges in a wide region of the phase diagram [in between the two magnetization plateaus of the triplon ``normal'' solid (NS) states]. The transition between the solid (NS or SS) and BEC states is of first order (thick lines). We show that the phase transition process upon cycling the magnetic field exhibits an anomalous magnetic hysteresis (see the schematic illustration in Fig.~\ref{fig1}) analogous to the melting transition in a triangular-lattice system of Bose gases~\cite{yamamoto-12}. In the anomalous hysteresis, although the transition from NS to BEC can occur (left panel), in the reverse process the NS and SS states are {\it ignored} and no reverse transition occurs (right panel). This phenomenon is completely different from a conventional first-order transition with a hysteresis loop. 

\section{\label{2}Spin dimers on a triangular lattice} 
% 5th
Below we will give a detailed description of the above-mentioned physics. 
The triangular-lattice spin-dimer compounds such as Ba$_3$Mn$_2$O$_8$~\cite{uchida-01,stone-08,samulon-09}, Sr$_3$Cr$_2$O$_8$~\cite{aczel-09,castro-10}, and Ba$_3$Cr$_2$O$_8$~\cite{kofu-09,aczel-09-2} have a stacked bilayer structure shown in the inset of Fig.~\ref{fig1}. In usual compounds, the coupling $J_1$ between the dimer planes is relatively weak compared to the inplane interactions $J_0$, $J_2$, and $J_3$~\cite{stone-08,castro-10,kofu-09}, and the number of $J_1$ bonds is only $1/6$ of the $J_2$ or $J_3$ one. Thus, we capture the essential physics of coupled spin dimers using the following frustrated $S=1/2$ spin-dimer model~\cite{sengupta-07,chen-10,albuquerque-11} on a single triangular-lattice plane:
\begin{eqnarray}
\hat{H}&=&
J_0\sum_{i}\hat{\mbox{\boldmath $S$}}_{i,1}\cdot\hat{\mbox{\boldmath $S$}}_{i,2}+J_2\sum_{\langle i,j\rangle}\left(\hat{\mbox{\boldmath $S$}}_{i,1}\cdot\hat{\mbox{\boldmath $S$}}_{j,1}+\hat{\mbox{\boldmath $S$}}_{i,2}\cdot\hat{\mbox{\boldmath $S$}}_{j,2}\right)\nonumber\\
&&+J_3\sum_{\langle i,j\rangle}\left(\hat{\mbox{\boldmath $S$}}_{i,1}\cdot\hat{\mbox{\boldmath $S$}}_{j,2}+\hat{\mbox{\boldmath $S$}}_{i,2}\cdot\hat{\mbox{\boldmath $S$}}_{j,1}\right)-H\sum_{i,\alpha}\hat{S}^z_{i,\alpha},
\label{hamiltonian}
\end{eqnarray}
where $\hat{\mbox{\boldmath $S$}}_{i,\alpha}=(\hat{S}^x_{i,\alpha},\hat{S}^y_{i,\alpha},\hat{S}^z_{i,\alpha})$ is the spin-1/2 operator attached to the site $i$, the two spins within a dimer are labeled by the subscript $\alpha=1,2$, and the second and third sums run over nearest-neighbor (NN) sites. Here $J_0$, $J_2$, and $J_3$ are the intradimer, $direct$ interdimer, and $crossed$ interdimer interactions, respectively (see the inset of Fig.~\ref{fig1}). We assume that the dimers form a regular triangular lattice and all the inplane couplings $J_0$, $J_2$, and $J_3$ are antiferromagnetic and isotropic in spin space as in a typical spin-dimer compound Ba$_3$Mn$_2$O$_8$~\cite{uchida-01,stone-08,samulon-09}. The last term is the Zeeman coupling with a magnetic field $H=g\mu_{\rm B}h$ (in energy units).

%%%%%%%%%%%%%%%%%%%%%%%%%%%%%%%%%%%%%%%%%%%%%%%%%%%%%%%%%%%%%%%%%%%%%%%%%%%%%%%
%%%%%%%%%%%%%%%%%%%%%%%%%%%%%%%%%%%%%%%%%%%%%%%%%%%%%%%%%%%%%%%%%%%%%%%%%%%%%%%
%%                                                                           %%
%% Section II: method                                                        %%
%%                                                                           %%
%%%%%%%%%%%%%%%%%%%%%%%%%%%%%%%%%%%%%%%%%%%%%%%%%%%%%%%%%%%%%%%%%%%%%%%%%%%%%%%
%%%%%%%%%%%%%%%%%%%%%%%%%%%%%%%%%%%%%%%%%%%%%%%%%%%%%%%%%%%%%%%%%%%%%%%%%%%%%%%
%
%%
\begin{figure}[b]
\includegraphics[scale=0.31]{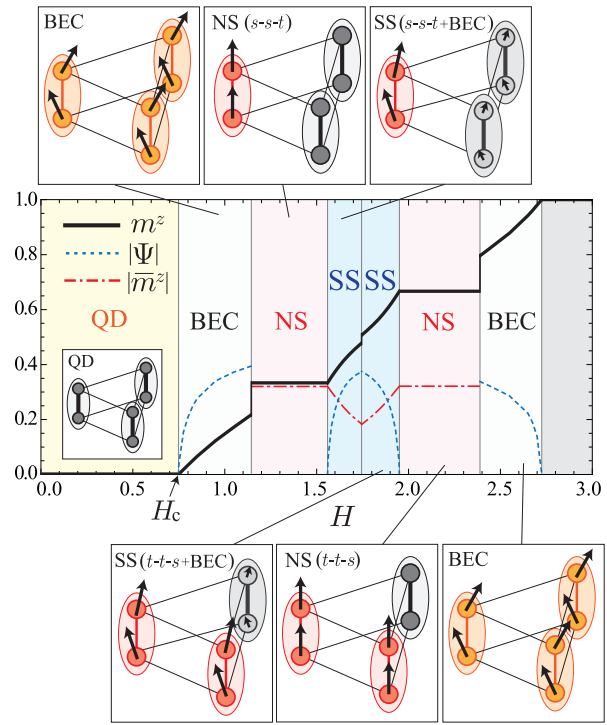}
\caption{\label{fig2} (Color online) 
The magnetization process at $J_2-J_3=-0.075$ with the illustrations of each phase. $\Psi$ and $\bar{m}^z$ are the order parameters of BEC and NS phases, respectively. 
}
\end{figure}
In the limit of $J_0,H\gg J_2,J_3$, the above model is mapped onto the hardcore Bose-Hubbard model with NN hopping $t=-(J_2-J_3)/2$ and NN repulsion $V=(J_2+J_3)/2$~\cite{mila-98}, which describes hardcore triplons moving in the singlet sea. We focus here on the case of $t\geq 0$ ($J_2\leq J_3$) as in the usual bosonic models in order to extract the Bose-Hubbard physics in the isotropic spin system given in Eq.~(\ref{hamiltonian}). However, the value of $J_0$ in the usual spin-dimer compounds is just one digit larger than $J_2$ or $J_3$~\cite{stone-08,castro-10,kofu-09}, and thereby one cannot regard the situation as the limit of $J_0\gg J_2,J_3$ while also giving consideration to the difference of coordination numbers between the inter- and intra-dimer couplings. 
Therefore, we study the ground-state properties of the spin-dimer model in the original form [Eq.~(\ref{hamiltonian})] for a realistic strength of the intradimer interaction, $J_2+J_3=0.5J_0$, away from the strongly dimerized limit. We set $J_0=1$ hereafter. Our CMF method~\cite{yamamoto-12-2} can take into account the effects of quantum and thermal fluctuations by exactly diagonalizing the cluster Hamiltonian with a mean-field boundary condition. Here we use the triangular-shaped clusters of $N_{\rm C}=1$, $3$, $6$, and $9$ dimers (2, 6, 12, and 18 spins) and then perform the cluster-size scaling. The CMF+S approach can treat quantitatively frustrated systems avoiding the notorious minus-sign problem. 

%%%%%%%%%%%%%%%%%%%%%%%%%%%%%%%%%%%%%%%%%%%%%%%%%%%%%%%%%%%%%%%%%%%%%%%%%%%%%%%
%%%%%%%%%%%%%%%%%%%%%%%%%%%%%%%%%%%%%%%%%%%%%%%%%%%%%%%%%%%%%%%%%%%%%%%%%%%%%%%
%%                                                                           %%
%% Section III: magnetic-field dependence                                    %%
%%                                                                           %%
%%%%%%%%%%%%%%%%%%%%%%%%%%%%%%%%%%%%%%%%%%%%%%%%%%%%%%%%%%%%%%%%%%%%%%%%%%%%%%%
%%%%%%%%%%%%%%%%%%%%%%%%%%%%%%%%%%%%%%%%%%%%%%%%%%%%%%%%%%%%%%%%%%%%%%%%%%%%%%%
% 6th
\section{\label{3}Magnetization process} 
In Fig.~\ref{fig1}, we already presented the ground-state phase diagram in the ($|J_2-J_3|,H$)-plane obtained by the CMF calculation with the nine-site cluster (CMF-9). 
To explain each phase in the phase diagram, we plot in Fig.~\ref{fig2} the magnetization curve $m^z(H)= \sum_i \langle \hat{S}_{i,1}^z+\hat{S}_{i,2}^z\rangle/M$ at $J_2-J_3=-0.075$ with the BEC ($\Psi$) and solid ($\bar{m}^z$) order parameters. Here $M$ denotes the number of dimers. At zero magnetic field, the antiferromagnetic intradimer interaction $J_0$ produces a gapped QD state with $m^z=0$, in which the spins in each dimer form a singlet state. This state cannot be described by a classical spin picture (or by the single-spin mean-field approximation). As the magnetic field $H$ increases, the energy of the triplet excitations with $S^z=+1$ gets lower and the system undergoes a transition into a magnetic state with closing the spin gap at the critical field $H_{\rm c}$. This phase transition can be understood as the BEC of magnetic quasiparticles (triplons)~\cite{nikuni-00,giamarchi-08}. The order parameter of the magnon BEC phase can be written as $\Psi=\sum_i\langle \hat{b}_i\rangle/M$ using the semihardcore boson operator $\hat{b}_i=(\hat{S}_{i,1}^+-\hat{S}_{i,2}^+)/\sqrt{2}$. In the original spin language, the BEC of triplons is translated as the canted antiferromagnetic state shown schematically in Fig.~\ref{fig2}. When the magnetic field increases further, the density of triplons gets closer to the commensurate filling of the triangular lattice, leading to the transition into the NS phase. The NS state at the magnetization plateau with $m^z=1/3$ has a $\sqrt{3}\times\sqrt{3}$ solid order with the ``singlet-singlet-triplet'' ($s$-$s$-$t$) superlattice pattern (Fig.~\ref{fig2}, upper middle panel), in which the original translational symmetry of the triangular lattice is broken. This state is characterized by the order parameter $\bar{m}^z=\sum_i\langle \hat{S}_{i,1}^z+\hat{S}_{i,2}^z \rangle \exp (i{\bf Q}\cdot {\bf r}_i)/M $ with ${\bf Q}=(4\pi/3,0)$. Additionally, we can see another plateau at $m^z=2/3$, in which the NS state has the triplet-triplet-singlet ($t$-$t$-$s$) pattern (Fig.~\ref{fig2}, lower middle panel). At higher fields, the system enters into the BEC phase again and eventually all the spins are aligned parallel to the field direction. 
.

\begin{figure}[t]
\includegraphics[scale=0.45]{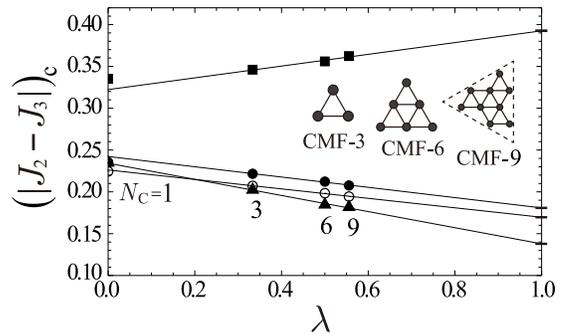}
\caption{\label{fig3} Examples of the cluster-size scalings of the phase boundaries between BEC-SS (triangle, $H=1.7$), BEC-NS (open circle, $H=1.4$; filled circle, $H=2.2$), and BEC-QD (square, $H=0$).} 
\end{figure}
%%
%
% 7th
In the region sandwiched between the two plateaus, both the BEC and solid order parameters can have a finite value. When applying a magnetic field (or doping triplons) on the $m^z=1/3$ NS state, the BEC of triplons occurs in the $s$-$s$-$t$ background and the magnon SS state emerges. 
The spins on both sublattice sites cant slightly away from the field direction (Fig.~\ref{fig2}, upper right panel). Similarly, another magnon SS state also emerges by doping holes of triplons on the $m^z=2/3$ NS state. As seen in the phase diagram of Fig.~\ref{fig1}, the magnon SS states can emerge over a reasonably large region in between the two NS states in contrast to the case of the square lattice~\cite{chen-10,sengupta-07,albuquerque-11,ng-06}. We also apply a cluster-size scaling (CMF+S) of the phase boundaries with the scaling factor $\lambda=\frac{N_{\rm B}}{3N_{\rm C}}$~\cite{yamamoto-12-2}, where $N_{\rm B}$ is the number of inter-dimer bonds treated exactly within the cluster. The denominator $3N_{\rm C}$ is the number of inter-dimer bonds per $N_{\rm C}$ sites in the original lattice. The parameter $\lambda$, which depends on both the number of cluster sites and the shape of the cluster, provides an indication of how much the correlation effects between the dimers are taken into account in the CMF calculation. Cluster approximations with mean-field boundary condition take into account the effects of the remaining spins on the infinite-size lattice via mean fields acting on the cluster spins. Thus the results usually converge much faster as the cluster size increases than finite-size calculations~\cite{yamamoto-12-2,yamamoto-12-3,luhmann-13}. In fact, the cluster-size scaling with $\lambda$ has successfully produced quantitatively reliable phase diagrams for related hardcore boson models~\cite{yamamoto-12-2,yamamoto-12-3}. We perform a linear fit of the data for the phase boundaries obtained with $N_{\rm C}=3$, $6$, and $9$ ($\lambda=1/3$, $1/2$, and $5/9$), which shows a good fit as shown in Fig.~\ref{fig3}. The scaled result denoted by the dashed green lines in Fig.~\ref{fig1} indicates that the SS region survives in the limit of $N_{\rm C}\rightarrow\infty$ ($\lambda\rightarrow 1$).

For the square lattice, the emergence of the magnon SS phase in the spin-dimer model [Eq.~(\ref{hamiltonian})] has been discussed with a numerical approach based on the tensor product states~\cite{chen-10}. The parameter region treated in Ref.~\onlinecite{chen-10} is $J_2+J_3\approx 0.5J_0$ as in the present work. 
The big difference from our result is that the SS phase is found only in a narrow region just below the $m^z=1/2$ plateau for spin dimers on the square lattice. On the other hand, the SS phase occupies a much wider region between the two magnetization plateaus in the present case of the triangular-lattice system. 
From the CMF+S phase boundaries in Fig.~\ref{fig1}, the maximum width of the supersolid region in the magnetic-field axis, $\delta H_{\rm SS}$, is evaluated to be about 15\% of the saturation field $H_{\rm s}$, which is approximately four times larger than the one for the square-lattice case~\cite{chen-10,sengupta-07,albuquerque-11}. The typical value of $H_{\rm s}$~\cite{uchida-01,stone-08,samulon-09,aczel-09,castro-10} gives $\delta H_{\rm SS}\sim$ 8 T, which can be clearly detectable in current experimental techniques.

%%%%%%%%%%%%%%%%%%%%%%%%%%%%%%%%%%%%%%%%%%%%%%%%%%%%%%%%%%%%%%%%%%%%%%%%%%%%%%%
%%%%%%%%%%%%%%%%%%%%%%%%%%%%%%%%%%%%%%%%%%%%%%%%%%%%%%%%%%%%%%%%%%%%%%%%%%%%%%%
%%                                                                           %%
%% Section IV: anomalous hysteresis                                          %%
%%                                                                           %%
%%%%%%%%%%%%%%%%%%%%%%%%%%%%%%%%%%%%%%%%%%%%%%%%%%%%%%%%%%%%%%%%%%%%%%%%%%%%%%%
%%%%%%%%%%%%%%%%%%%%%%%%%%%%%%%%%%%%%%%%%%%%%%%%%%%%%%%%%%%%%%%%%%%%%%%%%%%%%%%
\section{\label{4}Anomalous hysteresis} 
\begin{figure}[t]
\includegraphics[scale=0.5]{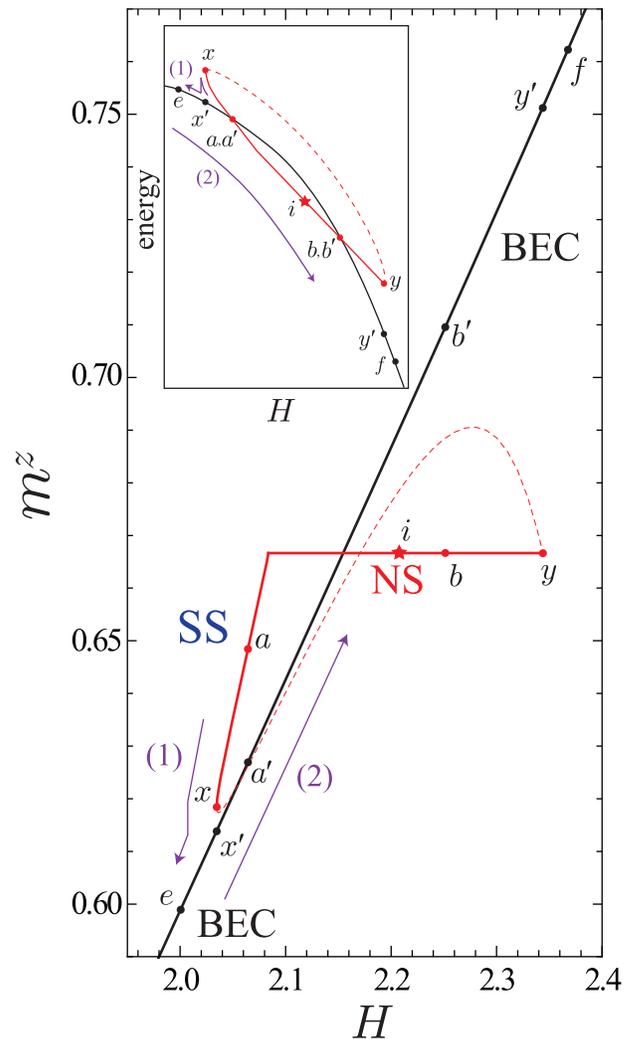}
\caption{\label{fig4} (Color online) Anomalous hysteresis behavior at $J_2-J_3=-0.101$. The solid and dashed lines correspond to the (meta)stable and unstable states, respectively. The inset is the schematic illustration of the characteristic structure of the energy. }
\end{figure}
%%
%

% 8th
Now we present a detailed description of the anomalous hysteresis behavior already mentioned earlier (Fig.~\ref{fig1}, lower panels). We see in the phase diagram that re-entrant (two-step) first-order transitions from BEC to solids (SS and NS), and back to BEC can be induced by a magnetic field near the tips of the NS phases. However, since the actual transition process in first-order phase transitions is determined not only by the ground states, we need to take into account also the metastable states. Figure~\ref{fig4} shows the CMF-9 data of the magnetization curve $m^z(H)$, including metastable and unstable solutions, around the $m^z=2/3$ plateau at $J_2-J_3=-0.101$. The stable (solid curve) and unstable (dashed curve) states correspond to the minima and maxima (or saddle points) of the energy landscape, respectively. When starting from an initial NS state, e.g. at point $i$, and decreasing the magnetic field, a transition to BEC occurs at the metastability limit of SS phase through the path $i\rightarrow a\rightarrow x\rightarrow x'\rightarrow e$. In the usual hysteresis, if we perform the reverse process (increase the field strength) starting from the BEC state, the system returns to the initial NS state through a different path, and a hysteresis loop is formed. However, surprisingly, the BEC state keeps the metastability even if increasing $H$ as clearly seen in the figure ($e\rightarrow f$). This unidirectional behavior is due to the absence of the metastability limit of the magnon BEC state. In this hysteresis cycle, obviously, a standard hysteresis loop is not formed and the Maxwell equal-area rule is not applicable. Although the energy of the system cannot be calculated directly in the large-size CMF~\cite{yamamoto-12-2}, the shape of the energy curve can be predicted by the magnetization curve $m^z(H)$. 
The curve of the energy should form the unconventional structure shown schematically in the inset of Fig.~\ref{fig4}, in which the line of BEC state passes through the loop consisting of NS and SS, instead of the usual {\it swallowtail} one~\cite{swallowtail}. Note that the same unidirectional behavior can be also obtained by increasing first and then decreasing $H$ (through the pass $i\rightarrow b\rightarrow y\rightarrow y'\rightarrow f$ and then $f\rightarrow e$) around the NS plateau with $m^z=1/3$. 

%%%%%%%%%%%%%%%%%%%%%%%%%%%%%%%%%%%%%%%%%%%%%%%%%%%%%%%%%%%%%%%%%%%%%%%%%%%%%%%
%%%%%%%%%%%%%%%%%%%%%%%%%%%%%%%%%%%%%%%%%%%%%%%%%%%%%%%%%%%%%%%%%%%%%%%%%%%%%%%
%%                                                                           %%
%% Section IV: extent of anomalous hysteresis                                %%
%%                                                                           %%
%%%%%%%%%%%%%%%%%%%%%%%%%%%%%%%%%%%%%%%%%%%%%%%%%%%%%%%%%%%%%%%%%%%%%%%%%%%%%%%
%%%%%%%%%%%%%%%%%%%%%%%%%%%%%%%%%%%%%%%%%%%%%%%%%%%%%%%%%%%%%%%%%%%%%%%%%%%%%%%

% 9th
%
%%
\begin{figure}[b]
\includegraphics[scale=0.35]{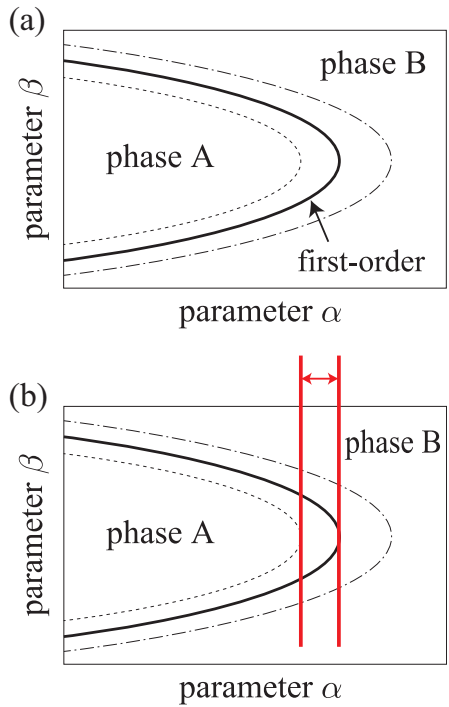}
\caption{\label{figS1} (Color online) (a) Typical phase diagram of a system which exhibits the anomalous hysteresis. The solid line is the first-order transition line. The dashed (dash-dotted) line corresponds to the metastability limit of phase B (phase A). (b) The range of parameter $\alpha$ for observing the anomalous hysteresis upon sweeping parameter $\beta$. }
\end{figure}
The unidirectional behavior in the anomalous hysteresis stems from only the characteristic geometry of the phase diagram. 
In order to generalize the discussion, let us first consider a simpler situation where a certain system has a phase diagram with the geometry shown in Fig.~\ref{figS1}(a). 
There is a first-order phase transition between phases A and B in the parameter $\alpha$ versus $\beta$ plane. The essential point is that phase A of lobe shape is completely surrounded by the first-order transition boundary. In this case, the metastability limit of phase B (dashed curve) is necessarily located inside the lobe, while the one of phase A (dash-dotted curve) is outside. As a result, there must be a finite region where phase B is always stable for any value of parameter $\beta$ [see the region indicated by the double-headed arrow in Fig.~\ref{figS1}(b)]. Therefore, when sweeping the value of $\beta$ at a fixed $\alpha$ in the region, one never encounters the metastability limit (spinodal) of phase B. Thus, the geometrical feature of the phase diagram leads to the unidirectional behavior in the anomalous hysteresis; whereas the transition occurs from phase A to B at the metastability limit of phase A (dash-dotted curve), a state in phase B is not dynamically destabilized due to the absence of spinodals of phase B along the sweeping path.

\begin{figure}[t]
\includegraphics[scale=0.3]{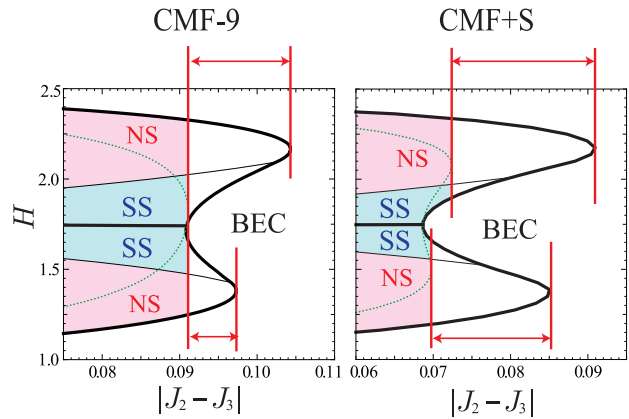}
\caption{\label{fig5} (Color online) The emergence region of the anomalous hysteresis behavior. The left and right panels show the CMF-9 and CMF+S data, respectively. The green dotted curves indicate the metastability limits of the BEC phase. The value of $|J_2-J_3|$ is required to lie within the range sandwiched between the two vertical red lines for the anomalous hysteresis.  }
\end{figure}
As shown in Fig.~\ref{fig5}, the ground-state phase diagram of the present spin-dimer model is more complicated but has the same geometry; the NS lobes are surrounded by first-order boundary and the metastability-limit curve of BEC phase (green dotted lines) is located inside the lobes. As a result, one never encounters the metastability limit of the BEC phase when sweeping the magnetic field in the region sandwiched between the two vertical red lines, which leads to the unidirectional transition process. In this case, there are two NS lobes at $m^z=1/3$ and $2/3$ and we have an opportunity to observe the anomalous hysteresis around the tip of either NS lobe. The required range of $|J_2-J_3|$ is different for each lobe (see the double-headed arrows in Fig.~\ref{fig5}). 
The blank (uncolored) regions of SS and NS are {\it ignored} upon sweeping the magnetic field from a BEC state. We have to use thermal cycling in order to reach these solid states. This means, at the same time, that the SS and NS states could be missed in experiments varying only the field strength at a fixed low temperature. For observing the anomalous hysteresis, the initial NS state could be prepared e.g., by cooling the sample in a high magnetic field.

%%%%%%%%%%%%%%%%%%%%%%%%%%%%%%%%%%%%%%%%%%%%%%%%%%%%%%%%%%%%%%%%%%%%%%%%%%%%%%%
%%%%%%%%%%%%%%%%%%%%%%%%%%%%%%%%%%%%%%%%%%%%%%%%%%%%%%%%%%%%%%%%%%%%%%%%%%%%%%%
%%                                                                           %%
%% Section V: summary                                                       %%
%%                                                                           %%
%%%%%%%%%%%%%%%%%%%%%%%%%%%%%%%%%%%%%%%%%%%%%%%%%%%%%%%%%%%%%%%%%%%%%%%%%%%%%%%
%%%%%%%%%%%%%%%%%%%%%%%%%%%%%%%%%%%%%%%%%%%%%%%%%%%%%%%%%%%%%%%%%%%%%%%%%%%%%%%
\section{\label{5}Summary} 
% 10th
In summary, we have used a large-size cluster mean-field theory in combination with a scaling scheme to study the ground-state phase diagram and the hysteresis of the frustrated $S=1/2$ spin-dimer model on a triangular lattice away from the strongly dimerized limit for $J_3\geq J_2$.
The magnetization curve exhibits two plateaus at $m^z=1/3$ and $2/3$, in which the singlet and triplet states form a $\sqrt{3}\times\sqrt{3}$ solid order with ordering vector ${\bf Q}=(4\pi/3,0)$ (the NS state of triplons). We found that more exotic magnon SS states emerge in between the two plateaus of the NS states. Since the SS phases occupy a broad region of the phase diagram in contrast to the square-lattice case, it should be easier to experimentally discover these phases in triangular spin-dimer materials.
Moreover, we found that the metastability of the system yields an anomalous hysteresis behavior without forming a standard hysteresis-loop structure upon cycling the magnetic field. In the anomalous hysteresis cycle, the actual transition can occur only unidirectionally from the solid (SS or NS) phase to the uniform magnon BEC phase. 
The Ginzburg-Landau description of the anomalous hysteresis will be reported elsewhere~\cite{yamamoto-12-3}.

%%%%%%%%%%%%%%%%%%%%%%%%%%%%%%%%%%%%%%%%%%%%%%%%%%%%%%%%%%%%%%%%%%%%%%%%%%%%%%%
%%%%%%%%%%%%%%%%%%%%%%%%%%%%%%%%%%%%%%%%%%%%%%%%%%%%%%%%%%%%%%%%%%%%%%%%%%%%%%%
%%                                                                           %%
%% Section VI: acknowledgements                                              %%
%%                                                                           %%
%%%%%%%%%%%%%%%%%%%%%%%%%%%%%%%%%%%%%%%%%%%%%%%%%%%%%%%%%%%%%%%%%%%%%%%%%%%%%%%
%%%%%%%%%%%%%%%%%%%%%%%%%%%%%%%%%%%%%%%%%%%%%%%%%%%%%%%%%%%%%%%%%%%%%%%%%%%%%%%
The authors thank Carlos A. R. S\'a de Melo for useful discussions. This work was supported by KAKENHI (23840054) from JSPS (D. Y.).

\end{document}